\documentclass{iau-JD}
\usepackage{graphicx}

\def\spose#1{\hbox to 0pt{#1\hss}}
\def\lta{\mathrel{\spose{\lower 3pt\hbox{$\mathchar"218$}}
     \raise 2.0pt\hbox{$\mathchar"13C$}}}
\def\gta{\mathrel{\spose{\lower 3pt\hbox{$\mathchar"218$}}
     \raise 2.0pt\hbox{$\mathchar"13E$}}}

{\newif\ifnotend
\notendtrue
\def\veclist{ABCDEFGHIJKLMNOPQRSTUVWXYZabcdefghijklmnopqrstuvwxyz.}
\def\top#1#2.{#1}
\def\tail#1#2.{#2.}
\loop\expandafter\xdef\csname v\expandafter\top\veclist\endcsname%
{{\noexpand\bf\expandafter\top\veclist}}
\edef\veclist{\expandafter\tail\veclist}
\if\veclist.\notendfalse\fi\ifnotend\repeat}
\pubyear{2009}
\volume{Volume 15}  %% insert here IAU Highlights of Astronomy Volume Number
\pagerange{1--30}
\date{?? and in revised form ??}
\jname{Highlights of Astronomy, Volume 15}
\editors{Ian F Corbett, ed.}
\begin{document}
 \title[Schwarzschild Models]{Schwarzschild Models for the Galaxy}

\author[Chanam\'e]{Julio Chanam\'e}

\affiliation{Carnegie Institution of Washington, Department of Terrestrial Magnetism \break 5241 Broad Branch Rd., Washington DC 20015, USA \break email: jchaname@dtm.ciw.edu}

\maketitle
\begin{abstract}

  Schwarzschild's orbit-superposition technique is the most
  developed and well-tested method available for constraining the
  detailed mass distributions of equilibrium stellar systems.  Here I
  provide a very short overview of the method and its existing
  implementations, and briefly discuss their viability as a tool for
  modeling the Galaxy using Gaia data.

%\keywords{Keyword1, keyword2, keyword3, etc.}
\end{abstract}

\firstsection % if your document starts with a section,
\section{Introduction}

Models are used to relate observations to theoretical constructs such as the
phase-space distribution function. A number of simplifying assumptions are
required to make the modeling process tractable and these assumptions can
have profound implications for the inferred mass distribution.  A delicate
balance must be struck between the available observations and the complexity
of the models fitted to them.

The Gaia mission will require modeling techniques that are capable of
handling huge numbers of measurements, while taking full advantage of the
high precision of the data.  Existing implementations of the
orbit-superposition method already fulfil some of these requirements but do
have limitations.  For example, while an assumption regarding the
geometry of the gravitational potential is inescapable, Schwarzschild models
can be built that are completely free from assumptions regarding the detailed
orbital structure, which generally have the largest impact on the derived
mass distribution.

\subsection{Schwarzschild's technique}

Given an orbit library for an assumed gravitational potential, the
orbit-superposition technique \cite[(Schwarzschild 1979)]{sch79} finds the linear sum of those
orbits that best reproduces the available observations.  The success of the
method relies on two aspects: (1) that the stellar system can be safely
considered to be in equilibrium, and (2) that the orbit library is
sufficiently comprehensive.  If these two conditions are satisfied, the
method is very general and free from most assumptions.  Even the required
assumption of a given geometry for the gravitational potential is in practice removed by the iterative
nature of the technique, which calls for the construction of Schwarzschild
models for an entire grid of potentials, with the final model the one that
best fits the data.  Of course a Schwarzschild model only provides a snapshot
of the current dynamical state of the system, and the question of stability
must be addressed by other means.

Schwarzschild models have been successfully used to constrain the dark-matter
halos of galaxies (e.g., Rix et al. 1997, Thomas et al. 2005), to weigh
supermassive black holes at the centers of both galaxies (e.g., van der Marel
et al. 1998, Gebhardt et al. 2000, 2001) and globular clusters (Gebhardt,
Rich, \& Ho 2005). They have also been used to study the dynamics of star
clusters (e.g., van de Ven et al. 2006).  More relevant to the subject at
hand, orbit-superposition models have also been used to study the dynamics of
the Galactic bulge (Zhao 1996, H$\ddot{\rm a}$fner et al. 2000).  Existing
implementations of the Schwarzschild method are usually classified/labeled
according to the geometry of the stellar systems they can be applied to
(spherical, axisymmetric, triaxial), and to the type of dataset they are
designed to handle (continuous or discrete; see Chanam\'e, Kleyna, \& van der
Marel 2008 for a review).

Points of weakness or controversy regarding Schwarzschild modeling include:
the non-uniqueness of the initial conditions used to generate orbits; the
amount of smoothing or regularization of the solution that is applied and its
impact on the final results; possible over-interpretation of $\chi^2$ plots
and indeterminacy of best solution; how to deal with incomplete positional
sampling; and large computational costs.

\section{Applicability of Schwarzschild's method to Galactic surveys}

We can consider the Galaxy to be composed of two kinds of structure: (1) a
smoothly distributed and old Galaxy in steady-state equilibrium, and (2) a
perturbed, inhomogeneous Galaxy that changes over relatively short timescales
and is not in dynamical equilibrium.  While the classical Galactic
structures of bulge, disk(s), and halo belong to the first category,
shorter-lived structures such as tidal streams, spiral arms, and disk warps,
all fall into the second one.  Only the background, steady-state Galaxy is
susceptible to Schwarzschild modeling. Fortunately, most of the Galaxy's mass
lies in steady-state structures, so a Schwarzschild model should provide a
useful first approximation to the data.  However, even though the
non-equilibrium mass fraction is small, this component is expected to hold
clues to the history of the Galaxy, and means must be found to model it too.

Modeling data from current surveys such as SDSS and RAVE will prepare us for
modeling the vastly superior Gaia data.  Clever arguments such as those in
Smith, Evans, \& An (2009) can only benefit the applicability of
Schwarzschild's technique by narrowing down the range of possible shapes and
geometries of the underlying gravitational potential, and could even shed
light on the optimal choice of initial conditions for orbit integration.

\begin{acknowledgments}

I thank the American Astronomical Society and
the International Astronomical Union  for travel grants.
I acknowledge support from NASA through Hubble Fellowship grant
HF-01216.01-A, awarded by the Space Telescope Science Institute,
which is operated by the AURA, Inc., under NASA contract NAS5-26555.

\end{acknowledgments}

\vfill\eject
\end{document}